\documentclass[aps,prl,twocolumn]{revtex4}

\begin{document}

\title{Anomalous fluctuations in phases with a broken continuous symmetry}

\author{W. Zwerger}

\affiliation{Institute for Theoretical Physics, University of Innsbruck, A-6020 Innsbruck, Austria}

\date{\today}

\begin{abstract}
  It is shown that the Goldstone modes associated with a broken continuous 
  symmetry lead to anomalously large fluctuations of the zero field order
  parameter at any temperature below $T_c$. In dimensions $2<d<4$,
  the variance of the extensive spontaneous magnetization scales as
  $L^4$ with the system size $L$, independent of the order parameter
  dynamics. The anomalous scaling is a consequence of the $1/q^{4-d}$
  divergence of the longitudinal susceptibility. For ground states in
  two dimensions with Goldstone modes vanishing linearly with momentum,
  the dynamical susceptibility contains a singular contribution
  $(q^2-\omega^2/c^2)^{-1/2}$.  The dynamic structure factor thus exhibits
  a critical continuum above the undamped spin wave pole, which may 
  be detected by neutron scattering in the N\'eel-phase of 2D quantum
  antiferromagnets.
 
\end{abstract}

\maketitle

It is one of the basic properties of any thermodynamic system that the 
fluctuations Var$\,\hat A=\langle\hat A^2\rangle-\langle\hat A\rangle^2\!\sim\! V$ of an extensive
variable $\hat A$ scale linearly with the system volume $V$. This 
property guarantees that the intensive variables $\hat A/V$ are 
self-averaging, with rms fluctuations vanishing proportional to $V^{-1/2}$.
Physically, the extensive nature of 
Var$\,\hat A\!\sim\! V$ is due to the existence of a finite, microscopic
correlation length $\xi$. A large system can thus be partitioned
into an extensive number of $V/\xi^d$ subvolumes which are
statistically independent. Since the variance in each subvolume 
is finite, the central limit theorem then quite generally implies
a Gaussian distribution for $\hat A$ with a variance of order $V$, 
in agreement with the standard Einstein theory of fluctuations
in macroscopic thermodynamics.

The above argument indicates that the linear scaling Var$\,\hat A\!\sim\! V$
may break down only at a critical point of a continuous phase 
transition where the correlation length $\xi$ diverges. In fact, from the 
standard relation Var$\,\hat M=V\cdot T\,\chi$ between the 
fluctuations of the total 'magnetization' $\hat M$ and the corresponding 
linear susceptibility, the finite size scaling $\chi(T_c,L)\sim L^{2-\eta}$
of the susceptibilty right at $T_c$ \cite{barber} implies a 
nontrivial dependence Var$\,\hat M\sim L^{d+2-\eta}$ on system size 
$L$ at the critical point.
Below, it will be shown that anomalous fluctuations of the order
parameter are present not only {\em at\/} $T_c$ but in fact at {\em \/any} 
temperature below  $T_c$ provided the broken symmetry is continuous.
This is a result  of the presence of Goldstone modes, which imply that
correlations decay algebraically below $T_c$ with exponents that are
independent of temperature. A specific example of recent interest
are  fluctuations of the condensate number $\hat N_0$ in a Bose Einstein
condensate. Using a weak coupling Bogoliubov approach, 
Giorgini et.al have shown \cite{giorgini} that Var$\,\hat N_0\sim T^2 L^4$
scale anomalously at low temperatures. In fact, this result also applies
to strongly interacting superfluids, being essentially a consequence of 
the long wavelength phase fluctuations \cite{meier}. In the present Letter, we will
show, that the anomalous fluctuations in a Bose-Einstein condensate are
just one particular example of a rather general phenomenon which
appears in any phase with a broken continuous symmetry. Consider, for
example, an isotropic ferromagnet in zero field below $T_c$. Because of
the invariance under spin rotations, it costs no energy to rotate the
direction of the magnetization vector. Provided that the microscopic 
interactions are short ranged, this implies that the transverse susceptibility
$\chi_{\perp}(q)$ diverges precisely as $1/q^2$ at small wavevectors $q\to 0$.
As a result, there are strong fluctuations of the {\em direction\/} of the
magnetization, leading to a complete destruction of long range order
at finite temperature
in dimensions $d\leq 2$, the well known Mermin-Wagner-Hohenberg 
theorem.  Regarding the {\em magnitude\/}
of the magnetization, the naive expectation is, that its fluctuations are just 
like that of a standard thermodynamic variable because there is a finite
restoring force for deviations from the equilibrium value. However, as was
noted a long time ago by Patashinski and Pokrovski \cite{patashinski},
the inevitable coupling between longitudinal and transverse order
parameter fluctuations entails that the longitudinal susceptibility is
also singular at small wave vectors, diverging as $\chi_{\|}(q)\sim 1/q^{4-d}$
in dimensions $2<d<4$.  Our aim in the following is to show that
\vspace*{0.2cm}
\hfill\newline
\hspace*{0.2cm}
a) the $1/q^{4-d}$ divergence of the longitudinal susceptibility
leads to anomalous fluctuations Var$\,\hat M_s\sim L^4$ of the zero field
order parameter in $2<d<4$ and $T\not= 0$ for an arbitrary broken 
continuous symmetry phase.  For superfluids, the corresponding {\em relative\/} 
condensate fluctuations are universal at low $T$.
\vspace*{0.2cm} 
\hfill\newline
\hspace*{0.2cm}
b) there is an analog of the singular nature of $\chi_{\|}$ for
zero temperature phases with a broken continuous symmetry. 
In two dimensions and with Goldstone modes whose 
frequencies $\omega=cq$ vanish linearly with momentum $q=|\vec q|$, the 
dynamical susceptibility has a singular contribution  $(q^2-\omega^2/c^2)^{-1/2}$
as noted first by Sachdev on the basis of a $1/N$ expansion \cite{sachdev}.
The dynamic structure factor of 2D quantum antiferromagnets
thus exhibits a critical continuum above the 
standard $\delta$-function spin wave peak. This effect may be observed in
high resolution neutron scattering experiments.

As an effective description of an arbitrary phase with a broken
continuous symmetry at finite temperature we use the nonlinear
$\sigma$-model \cite{parisi}. It describes directional
fluctuations of an order parameter $\vec\psi (x)=m_s^{(0)}\vec\Omega (x)$
with a fixed magnitude $m_s^{(0)}$ in terms of an $N$-component 
unit vector  $\vec\Omega (x)$. At zero external field, the effective action 
for the fluctuations of $\vec\Omega$ is
\begin{equation}
S\left[\Omega\right]=\frac{\rho_s}{2T}\,\int d^dx \left(\nabla\vec\Omega(x)\right)^2
\hspace*{0.5cm} |\vec\Omega(x)|^2=1, 
\end{equation}
with the spin stiffness (or helicity modulus) $\rho_s$ as the single phenomenological
parameter. In a finite system of volume $V=L^d$, there is of course no spontaneous
magnetization at zero field. The breaking of a continuous symmetry below
$T_c$ however shows up in the integrated $O(N)$-symmetric correlation
function at finite volume
\begin{equation}
\int_V d^dx\;\langle\vec\psi(x)\cdot\vec\psi(0)\rangle_L=Vm_L^2\to Vm_s^2,
\end{equation}
defining an intensive nonzero order parameter $m_L^2$ which approaches 
the spontaneous magnetization $m_s^2$ of the infinite system in the 
thermodynamic limit $L\to\infty$. For superfluids  this is just the number
of particles in the condensate. In dimensions $d>2$, where $m_s$
is nonzero at finite $T$, the leading long distance behaviour of the 
two point function $G(r)=\langle\vec\psi(x)\cdot\vec\psi(0)\rangle$ may be obtained 
from a simple Gaussian spin wave calculation. Following a standard
procedure \cite{parisi}, the unit vector $\vec\Omega(x)=\left( \vec\Pi(x), \sqrt{1-\vec\Pi(x)^2}
\right)$ is decomposed into $N-1$ 'transverse' Goldstone fields
$\Pi_a(x)\, , a=1, \ldots N-1$ and a longitudinal component
$\Omega_N(x)=\sqrt{1-\vec\Pi(x)^2}$. At low enough temperatures, where spin wave
interactions can be neglected, the $\Pi$ fields are Gaussian random variables
with variance $\langle\Pi_a(q)\,\Pi_b(q')\rangle=\delta_{a,b}\delta_{q,-q'}\, T/\rho_s q^2$. The
zero field correlation function below $T_c$
\begin{equation}
G(r)=m_s^2\left[ 1+ C_{\|}(r)+(N-1)\cdot C_{\perp}(r)\right]
\end{equation}
is thus split into longitudinal and transverse parts, with $m_s^2=G(\infty)$
the renormalized value of the spontaneous magnetization.
To lowest nontrivial order in the small fluctuations $\Pi$, the transverse
correlation function decays proportional to $T/\rho_sr^{d-2}$, as expected from the 
standard $1/q^2$ divergence of the transverse susceptibility 
\begin{equation}
\chi_{\perp}(q)=m_s^2 C_{\perp}(q)/T=\frac{m_s^2}{\rho_s q^2}
\end{equation}
in the symmetry broken phase. In leading order, the longitudinal function is
given by ($c$ stands for 'connected')
\begin{equation}
C_{\|}(r)\approx \frac{1}{4}\langle\vec\Pi^2(x)\vec\Pi^2(0)\rangle_c=\frac{N-1}{2} C_{\perp}^2(r)
\end{equation}
and thus again decays algebraically as $1/r^{2(d-2)}$. Contrary to the naive mean
field picture, where the londitudinal susceptibility $\chi_{\|}(q)$ below $T_c$
is finite as $q\to 0$, the slow decay of $C_{\|}(r)$ implies that 
$\chi_{\|}(q\to 0)\sim T/\rho_s^2q^{4-d}$ is divergent in $d<4$ \cite{patashinski}.
While the relation (5) is valid only to leading order in an expansion in powers
of $\Pi$, the behavior $C_{\|}(r)\sim r^{-2(d-2)}$ is in fact expected to be exact
at arbitrary temperatures below $T_c$ \cite{parisi}, consistent with a rigorous
correlation inequality for $N=2$ \cite{dunlop}.

To discuss the fluctuations of the spontaneous magnetization in a finite
system at zero field, we define a fluctuating, extensive variable $(d1=d^dx_1)$
\begin{equation}
\hat M_s=\frac{1}{V}\int\! d1\int\! d2\;\vec\psi(1)\cdot\vec\psi(2)
\end{equation}
with thermal average $\langle\hat M_s\rangle=Vm_L^2$. Its fluctuations are then
determined by the connected four point function
\begin{equation}
u_4(1234)=\langle\vec\psi(1)\cdot\vec\psi(2)\,\vec\psi(3)\cdot\vec\psi(4)\rangle_c\, .
\end{equation}
Expanding this consistently up to order $\Pi^4$, one finds that in an
infinite system this function may be expressed simply as
\begin{eqnarray}
\lefteqn{u_4(1234)=m_s^4\,\frac{N-1}{2} \cdot} \nonumber\\
& & \cdot\left[ C_{\perp}(r_{13})-C_{\perp}(r_{14})-C_{\perp}(r_{23})+C_{\perp}(r_{24})\right]^2\, .
\end{eqnarray}
In order to obtain the scaling of Var $\hat M_s$ in a finite system, we switch to a 
momentum representation and replace integrals $\int_q$ by discrete sums 
$V^{-1}\sum_q  \mbox{}^{\prime}$ over wavevectors. The $q=0$ contribution is excluded since it
describes an irrelevant global rotation of the order parameter $\vec\psi(x)$.  It is
then straightforward to show that
\begin{equation}
{\rm Var}\, \hat M_s=2m_s^4(N-1)\left(\frac{T}{\rho_s}\right)^2\cdot\sum_q  \mbox{}^{\prime}\frac{1}{q^4}
\end{equation}
which is proportional to $L^4$ in $2<d<4$ by simple dimensional analysis \cite{note1}.
Defining a numerical coefficient $B$ by $\sum_q \mbox{}^{\prime}q^{-4}=BL^4/8\pi^2$, we find 
that $B=8E_3(2)/\pi^2=0.501$ for a 3D box with Dirichlet boundary conditions.
Here $E_d(t)=\sum_{n_1=1, \ldots n_d=1}^{\infty}\, (n_1^2+\ldots  +n_d^2)^{-t}$
is the generalized Epstein zeta function, convergent for $d<2t$ (for periodic 
boundary conditions $B=0.8375$ \cite{meier}). 
 
Although our derivation of the general result $(9)$ is based on an expansion in 
powers of $\Pi$ and thus
appears to be restricted to low temperature, the exponent in Var $\hat M_s\sim L^4$
is universal below $T_c$ \cite{note2} just like the $q^{-(4-d)}$ 
divergence of the longitudinal susceptibility. Similarly, the temperature dependence  of
Var $\hat M_s$ will be $\left( T/\rho_s(T)\right)^2$ for arbitrary temperatures 
below $T_c$, vanishing proportional to $T^2$ at low temperatures since $\rho_s(T=0)$ is
finite. This follows from the fact that at any temperature $T$ in the broken symmetry
phase, the dominant finite size dependence is
determined by the leading low energy constant in the effective field theory for fluctuations
of the order parameter, which is precisely $\rho_s$ as defined in equation $(1)$. 
New effective constants only enter in higher order terms, as discussed for instance
in the context of chiral perturbation theory in QCD \cite{hasenfratz}.
From eqn.$(9)$, it follows that the relative fluctuations 
Var $\hat M_s$/$\langle\hat M_s\rangle^2$ scale as $L^{-2(d-2)}$. The spontaneous magnetization is
therefore self-averaging in $d>2$ although weaker than expected 
naively, unless $d>4$.
In this context, it should be mentioned that, 
in the case of a non-Abelian symmetry $N\geq 3$, the  
spin stiffness $\rho_s$ has anomalous fluctuations 
\begin{equation}
\frac{{\rm Var}\, \rho_s}{\langle\rho_s\rangle^2}\sim (N-2)\cdot\left(\frac{T}{\rho_sL^{(d-2)}}\right)^2
\end{equation}
on its own, as shown by Chakravarty \cite{chakravarty}. They are 
very similar to the fluctuations of the spontaneous magnetization
discussed here and indeed arise from the same kind of Goldstone anomalies. 
A special situation appears in homogeneous superfluids $(N=2)$
at low temperatures. Since translation invariance requires the superfluid 
density to be equal to the full density $n$, the associated stiffness
$\rho_s(T\to 0)=\hbar^2n/m$ is independent of the interaction. Equation $(9)$
then leads to the remarkable result that the relative fluctuations of the 
condensate number at low temperature $(d=3)$   
\begin{equation}
\lim_{T\to 0}\,\frac{{\rm Var}\, \hat N_0}{\langle\hat N_0\rangle^2}=
\frac{B}{\left(n\lambda_T^2L\right)^2}
\end{equation}
are completely universal, depending only on density, system size and the
thermal wavelength $\lambda_T=h/\sqrt{2\pi mT}$, but not on whether the 
superfluid is  weakly or strongly interacting \cite{remark}.
Finally, it is important to realize that - as in the case of the
Mermin-Wagner-Hohenberg theorem -  
the result $(9)$ is independent of the order parameter dynamics. 
It applies equally to say ferro- or antiferromagnets even though the 
temperature dependence of the {\it average} order parameter 
$\langle\hat M_s\rangle$ is, of course,
very different  in both cases. This may be shown either by a microscopic 
derivation of equation $(9)$ on the basis of noninteracting spin wave 
theory around a perfectly ordered ground state or, alternatively, by
using a quantum mechanical generalization of the nonlinear
$\sigma -$ model which describes magnons with dispersion
either $\omega\sim q^2$ or $\omega\sim q$.

As shown above, the anomalous fluctuations $(9)$ of the order
parameter are a consequence of the $q^{-(4-d)}$ divergence of the
longitudinal susceptibility in $d< 4$. In the following we want to discuss
the analog of this phenomenon
in zero temperature phases where a continuous symmetry is broken.
In this case the specific dynamics of the order parameter
is important.  Here, we assume that the Goldstone modes
have a linear spectrum $\omega=cq$ as in superfluids and antiferromagnets.
An effective description of the ordered phase can then be obtained
from a quantum mechanical nonlinear $\sigma-$model, with a unit
vector $\vec\Omega(x,\tau)$ which depends both on space $x$ and 
imaginary time $\tau\in [0,\beta\hbar]$. The corresponding effective 
action 
\begin{equation}
S\left[\Omega\right]=\frac{\rho_s}{2\hbar}\int_0^{\beta\hbar} d\tau\int d^dx \left[\left(\nabla\vec\Omega(x)\right)^2+\left(\frac{1}{c}\partial_{\tau}\vec\Omega\right)^2
\right] 
\end{equation}
has the spin wave velocity $c$ as the only additional parameter. Together with the
renormalized value $m_s^2$ of the long range order, $\rho_s$ and $c$ completely
determine the low energy properties of the ordered phase. As an example,
this model applies both to the N\'eel ordered state of 2D quantum antiferromagnets
discussed extensively in the context of high temperature superconductors \cite{chn}
or to the superfluid phase of cold atoms in an optical lattice which
has been realized recently in 3D \cite{greiner}.  Using again 
a lowest order expansion in powers
of the small fluctuations $\vec\Pi(x,\tau)$ in the standard decomposition 
$\vec\Omega=\left( \vec\Pi, \sqrt{1-\vec\Pi^2}\right)$, the transverse correlation
function at $T=0$ in two dimensions is given by \cite{note3}
\begin{equation}
C_{\perp}(x,\tau)=\frac{\hbar c}{4\pi\rho_s}\cdot\frac{1}{\sqrt{r^2+(c\tau)^2}}\, .
\end{equation}
Analytic continuation to real time $t=-i\tau$ and Fouriertransformation give
the standard form of the transverse dynamical susceptibility in any dimension
\begin{equation}
\chi_{\perp}(q,\omega)=\frac{m_s^2}{\rho_s}\cdot\frac{1}{q^2-\omega^2/c^2}\, .
\end{equation}
It leads to the expected quasiparticle pole at $\omega=cq$, reflecting the 
presence of undamped antiferromagnetic spin waves.
Considering the longitudinal correlations, the analog of the relation
$(5)$ again applies to leading order. As a result, the longitudinal
dynamical susceptibility in 2D turns out to be
\begin{equation}
\chi_{\|}(q,\omega)=\frac{(N-1)m_s^2\hbar c}{16\rho_s^2}\cdot
\frac{1}{\sqrt{q^2-\omega^2/c^2}}\, ,
\end{equation}
which has a branch cut rather than a simple pole. Formally
the result $(15)$ is completely analogous to the $1/q$ divergence
of the classical static susceptibility $\chi_{\|}(q)$ in $d=3$ discussed 
above. Indeed the effective dimensionality of the 2d quantum
antiferromagnet is $d+z=3$ and the dependence on frequency 
$\omega$ is dictated by the formal Lorentz-invariance of the 
model $(12)$.  

In order to relate these results to directly observable quantities, we
consider the rotationally averaged staggered susceptibility
(for antiferromagnets $N=3$)
\begin{equation}
\chi_{s}(q,\omega)=\frac{N-1}{N}\chi_{\perp}(q,\omega)+\frac{1}{N}\chi_{\|}(q,\omega)\, .
\end{equation}
For frequencies $\hbar\omega\gg T$, its imaginary part is equal to the 
dynamic structure factor $S(q,\omega)$ at $\omega>0$ up to a factor
$2\hbar$, giving
\begin{equation}
S(q,\omega)=2m_s^2\xi_J\frac{N-1}{N}\left[ \frac{\pi}{2q}\delta(\omega-cq)+
\frac{\xi_J}{16}\frac{\theta(\omega-cq)}{\sqrt{\omega^2-c^2q^2}}\right]\, .
\end{equation}
The longitudinal fluctuations of the N\'eel order thus lead to a critical
continuum above the spin wave pole at $\omega=cq$, which decays
only algebraically. The continuum results from the decay of a normally
massive amplitude mode with momentum $\vec p$ into a pair
of spin waves with momenta $\vec q$ and $\vec p -\vec q$, which 
is possible for any $\omega>cq$, with a singular cross section 
because of the large phase space. The amplitude mode is thus 
completely overdamped in two dimensions \cite{sachdev}. 
The relative weight of these fluctuations compared to
the dominant transverse contribution is determined by the 
dimensionless parameter $q\xi_J$, with $\xi_J=\hbar c/\rho_s$
the Josephson correlation length.  This length controls the 
crossover from the Goldstone regime $q\xi_J\ll 1$, where the spin dynamics
is well described by small fluctuations around a N\'eel ordered
ground state to quantum critical fluctuations at $q\xi_J\gg 1$. 
The quantum critical regime has 
a dynamical susceptibility of the form \cite{csy}
\begin{equation}
\chi_{s}(q,\omega)=\frac{m_s^2}{\rho_s}\left(\frac{N\xi_J}{2\pi}\right)^{\eta}
\cdot \frac{A_Q}{\left(q^2-\omega^2/c^2\right)^{1-\eta/2}}
\end{equation}
with an amplitude $A_Q$ close to one and an exponent $\eta$ 
which is nearly zero. Now, in the regime where the ground state 
exhibits strong N\'eel order, the Josephson correlation length
is only several lattice constants \cite{chn} and thus the
quantum critical regime is hardly accessible. In turn, there is
a rather wide Goldstone regime characterized by $q\xi_J\ll 1$
and $q\xi(T)\gg 1$, where $\xi(T)$ is  the finite 2d 
correlation length over which N\'eel order is lost at 
finite temperature.  Due to the exponential dependence
\cite{chn}
\begin{equation}
\xi(T)\sim\xi_J\exp{(2\pi\rho_s/T)}
\end{equation}
the relevant range of wavevectors $\xi^{-1}(T)<q<\xi_J^{-1}$ is rather wide at 
low temperatures. Experimentally,  spin waves in 2D quantum 
antiferromagnets have been observed by inelastic 
neutron scattering in the vicinity of the N\'eel ordering
wave vector $(\pi,\pi)$. In a constant $\omega$ scan, 
the first term in $(17)$ gives rise to a sharp peak at
$q=\omega/c$ with an amplitude $\sim 1/q$ \cite{aeppli}. 
The second contribution in $(17)$ due to the longitudinal
fluctuations implies an additional algebraic tail towards
smaller wave vectors, provided $q$ is in the 
Goldstone regime.  At room temperature,
where $\xi(T)$ is several hundred \AA\
and with typical values of  $\xi_J$ this gives 
a range $0.005<q$(\AA$^{-1})<0.05$.
Given the resolution in ref. \cite{aeppli}, detection of this 
algebraic tail appears very difficult, however high resolution
measurements may be able to observe the additional
contribution from the longitudinal spin fluctuations, which 
apparently behave like the critical fluctuations $(18)$,
however with a rather large exponent $\eta=1$.

In summary, it has been shown that Goldstone modes
associated with a broken continuous symmetry lead
to fluctuations of the zero field order parameter
which scale proportional to $L^4$ at any temperature 
below $T_c$. For 2D ordered ground states like the N\'eel 
phase of high temperature superconductors, the 
underlying longitudinal fluctuations lead to an additional
critical continuum beyond the simple spin wave pole
in the dynamic structure factor, whose detection requires
high resolution neutron scattering experiments.

I would like to thank the theory department 
at the ETH in Z\"urich, where this work was started, for providing  
a very stimulating environment.
Useful remarks by A. Aharony, R. Balian, D. Belitz, J. Fr\"ohlich, M. L\"uscher,
C. Newman and S. Sachdev are gratefully acknowledged.

\end{document}